\documentclass[%
 reprint,
superscriptaddress,
 amsmath,amssymb,
 aps,
]{revtex4-1}

\usepackage{graphicx}
\usepackage{dcolumn}
\usepackage{bm}
\usepackage{float}
\usepackage{color}

\begin{document}

\preprint{APS/123-QED}

\title{Self-assembly in soft matter with multiple length scales}

\author{Alberto Scacchi}\email{alberto.scacchi@aalto.fi}
\affiliation{Interdisciplinary Centre for Mathematical Modelling and Department of Mathematical Sciences, Loughborough University, Loughborough, Leicestershire LE11 3TU, UK}
\affiliation{Department of Chemistry and Materials Science, Aalto University, P.O. Box 16100, 00076 Aalto, Finland}
\affiliation{Department of Applied Physics, Aalto University, P.O. Box 11000, FI-00076 Aalto, Finland}

\author{Sousa Javan Nikkhah}
\affiliation{Department of Chemistry and Materials Science, Aalto University, P.O. Box 16100, 00076 Aalto, Finland}

\author{Maria Sammalkorpi}
\affiliation{Department of Chemistry and Materials Science, Aalto University, P.O. Box 16100, 00076 Aalto, Finland}
\affiliation{Department of Bioproducts and Biosystems, Aalto University, P.O. Box 16100, FI-00076 Aalto, Finland}

\author{Tapio Ala-Nissila} 
 \affiliation{Interdisciplinary Centre for Mathematical Modelling and Department of Mathematical Sciences, Loughborough University, Loughborough, Leicestershire LE11 3TU, UK}
 \affiliation{Quantum Technology Finland Center of Excellence and Department of Applied Physics, Aalto University, P.O. Box 11000, FI-00076 Aalto, Finland}

\date{\today}

\begin{abstract}
Spontaneous self-assembly in molecular systems is a fundamental route to both biological and engineered soft matter. Simple micellization, emulsion formation, and polymer mixing are well understood. However, the principles behind emergence of structures with competing length scales in soft matter systems remain unknown. Examples include droplet-inside-droplet assembly in many biomacromolecular systems undergoing liquid-liquid phase separation, analogous multiple emulsion formation in oil-surfactant-water formulations, and polymer core-shell particles with internal structure. We develop here a microscopic theoretical model based on effective interactions between the constituents of a soft matter system to explain self-organization both at single and multiple length scales. The model identifies how spatial ordering at multiple length scales emerges due to competing interactions between the system components, e.g. molecules of different sizes and different chemical properties. As an example of single and multiple-length-scale assembly, we map out a generic phase diagram for a solution with two solute species differing in their mutual and solvent interactions. We further connect the phase diagram to a molecular system via molecular simulations of a block-copolymer system that has a transition from regular single-core polymer particles to multi-core aggregates that exhibit multiple structural length scales.  The findings provide guidelines to understanding the length scales rising spontaneously in biological self-assembly, but also open new venues to the development and engineering of biomolecular and polymeric functional materials and pharmaceutical formulations.
\end{abstract}

\maketitle
%
Self-assembly is Nature's ingenious route to create new materials with complex structural and functional properties. Examples range from biological self-assembly, such as protein assemblies \cite{pieters2016}, cellular condensates \cite{banani2017biomolecular}, viruses \cite{zandi2020}, or cell membranes and their internal structure, such as lipid rafts \cite{Sezgin2017,meyers2008}, to widely used systems emerging from, e.g., polymer self-assembly \cite{stuart2010emerging,macfarlane2020,lu2020},  molecular crystals \cite{smalyukh2018,Phillips2016} or bioinspired approaches to materials engineering \cite{Gong2019,pieters2016}.
%

Particularly fascinating self-assembly materials are those exhibiting multiple length scales in their spatial arrangement. Examples include hierarchical biological and bioinspired materials, such as bone, nacre or crustacean exoskeletons \cite{meyers2008}, and silk-like materials \cite{wang2018}, all featuring exceptional toughness and resilience. Many block-copolymer systems \cite{lu2020}, coacervate droplets in biocondensates (liquid-liquid phase-separation in biological systems) \cite{banani2017biomolecular,lu2020multiphase} and multiple emulsions \cite{sheth2020multiple} readily exhibit hierarchical multiple length scales when self-assembling. This associates with locally varying molecular environments in terms of density and confinement, dielectric properties, or hydrophobicity and hydrophilicity, granting access to fascinating applications in nanophotonics \cite{ma2020,Phillips2016,Qi2019}, organic electronics \cite{casalini2017,Phillips2016}, confined catalysis \cite{York2017,Gaitzsch2016}, energy materials \cite{Gong2019} and sensors \cite{York2017,Phillips2016,Qi2019}.
Another important field of application is pharmaceutical materials and drug delivery \cite{York2017,lu2020,Qi2019}, where the heterogenous, often compartmentalized, solvation environment is interesting for sequential delivery of multiple drug species, for instance in cancer therapy \cite{Aw2013,Wu2018}.  

Simple molecular self-organization, such as aqueous micellization and emulsion formation, can be easily explained at the level of the interplay between water entropy, the relevant surface tensions, and the corresponding free energies \cite{nagarajan1991theory}. Multiple structural length scales emerge in the presence of competing interactions \cite{palermo2007,lu2020}, yet fundamental theoretical underpinning to deeply understand this is lacking.
We present here a theoretical framework demonstrating how competition between molecular interactions can lead to spontaneous formation of structurally complex self-assembly matter exhibiting  multiple-length-scale structural features and demonstrate its connection to a block-copolymer system.

We employ energy minimization principles based on classical density functional theory (DFT) \cite{hansen_mcdonald, evans_79, evans_92, lutsko2010recent} to model the spatially varying average density of each component in the thermodynamic equilibrium in a multi-component soft matter system. The interactions are approximated by a free-energy potential. This approach provides a generic phase diagram exhibiting a variety of possible equilibrium assembly configurations as a function of the system composition. We present a formulation that covers both the formation of single length-scale, simple phases and a multiple-periodic phase. The latter is commensurate with, e.g. formation of droplets inside droplets, such as biomolecular condensates \cite{banani2017biomolecular}, or multicore polymer micelles  \cite{
chen2013formation, duxin2005cadmium, 
ueda2011unicore, chen2012formation}.
%
\begin{figure}[t!]
\includegraphics[width=\columnwidth]{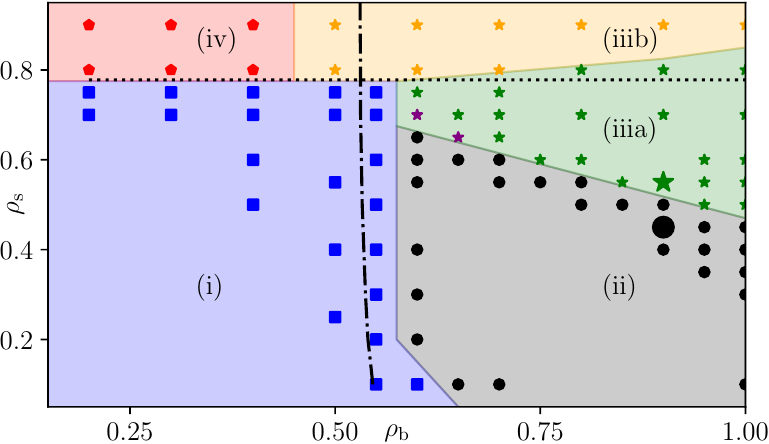}\caption{Generic assembly phase diagram of a 2D model binary system. Four distinct phases emerge: (i) a phase where both b and s particle densities are spatially uniform (squares), (ii) a phase corresponding to one species condensing into droplets surrounded by the other, i.e. a single-core phase (dots) (results for the large dot shown in Fig.~\ref{circle}), (iii) a multi-length-scale phase (stars) in which the species assemble into droplets inside droplets with the response corresponding to two subsets: (iiia) multi-core micelle type assembly (results for the large star shown in Fig.~\ref{star}) where each s island is clearly separate from the other s islands and (iiib) where the s islands overlap at their outskirt regions, and (iv) a phase where the s particles form a hexagonal lattice covering the whole space while the b particles are homogeneously spread (rhombi). This corresponds to species b being soluble while s forms droplets. Purple stars are hybrid cases, where most of the islands of s particles behave as the green star states. However, some random clusters do not display the short spacing and the distribution of the s particles is similar to the one of the b particles (see Fig.~\ref{star}(a)). The dotted line represents the linear instability of the length scale associated with the s particles, $k_{\rm s}$, and the dashed-dotted line the one associated with the b particles, $k_{\rm b}$. The boundary lines are guides to the eye.}\label{diagram}
\end{figure}
%

For computational efficiency, we consider a two-dimensional (2D) model system composed of two species in a solvent, however, the formalism is also valid in 3D. The system is characterized by soft interactions, where the effective pair potentials between coarse-grained complex molecules are designed using the generalized exponential model of index $n$ (GEM-$n$). Such potentials have been extensively used to describe the effective interactions in a vast variety of polymeric systems
\cite{bolhuis2001accurate, likos2001effective, 
likos1998star,louis2000can, 
gotze2004tunable, 
likos2006soft, lenz2012microscopically, mladek2006formation}. GEM-$n$ models exhibit interesting pattern formation if they are part of the $Q\pm$-interactions class~\cite{likos2007ultrasoft}, i.e. for $n>2$. This theory applies generally to any molecular system that has three partially immiscible components. One of the components can be considered as the solvent, implicitly present in the interaction potentials of the two explicit species. Such systems are common in e.g. aqueous polymer mixtures, biomolecular and colloidal systems, emulsions, and various liquid-crystal-forming systems.

The statistical distribution of the average densities $\rho_i(\textbf{r})$ are obtained using the Ramakrishnan-Yussouff approximation \cite{ramakrishnan1979first} (details in the Supplementary Material (SM)). 
The fundamental condition for thermal stability of any self-assembled structure is that it corresponds to a minimum in the relevant thermodynamic potential. We choose a mixture of by big (b) and small (s) coarse-grained particles interacting via the effective potentials
\begin{align}\label{pairpot}
\phi_{\rm bb}(r)&=\varepsilon_{\rm bb}e^{-\left(r/R_{\rm bb}\right)^4};\quad\quad \phi_{\rm ss}(r)=\varepsilon_{\rm ss}e^{-\left(r/R_{\rm ss} \right)^8}; \nonumber  \\
\phi_{\rm bs}(r)&=\varepsilon_{\rm bs}e^{-\left(r/R_{\rm bs} \right)^2}+\varepsilon_{\rm bs}^{+}e^{-\frac{1}{2}r/R_{\rm bs}},
\end{align}
\begin{figure*}[t!]
\includegraphics[width=1.0\linewidth]{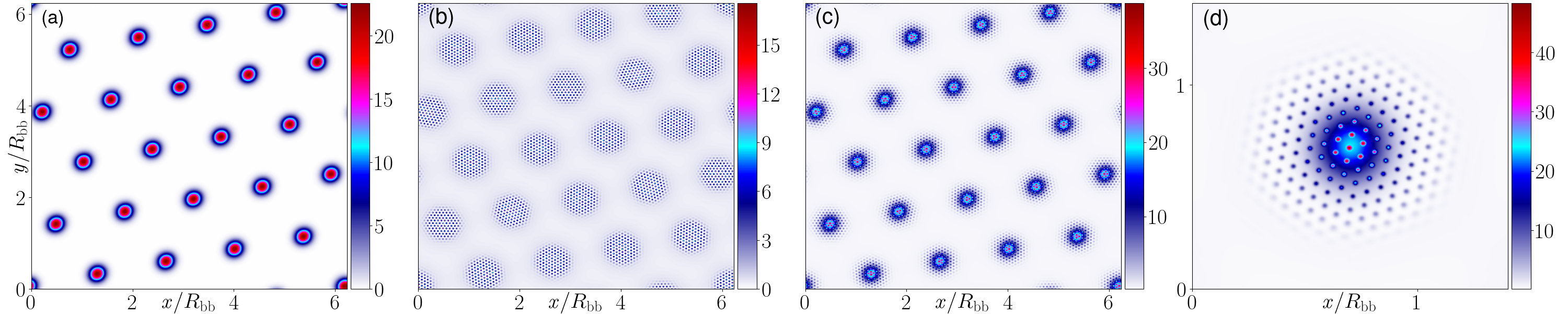}\caption{Dual-length-scale self-assembled phase (large star in Fig.~\ref{diagram}, $\rho_{\rm b}=0.9$, $\rho_{\rm s}=0.55$). For polymer systems, this refers to multi-core micelles, and for an emulsion to a thermodynamically stable droplet-inside-droplet state (multiple emulsion). (a) shows $\rho_{\rm b}(\textbf{r})$, (b) $\rho_{\rm s}(\textbf{r})$, (c) the total density $\rho_{\rm t}(\textbf{r})=\rho_{\rm b}(\textbf{r})+\rho_{\rm s}(\textbf{r})$, and (d) a blow-up of a single peak. $\rho_b(\textbf{r})$ exhibits hexagonal structure with spacing $\approx R_{\rm bb}$. Cocentric with $\rho_b(\textbf{r})$ peaks, also $\rho_s(\textbf{r})$ forms smaller hexagonal clusters with spacing $\approx R_{\rm ss}$.}\label{star}
\end{figure*}

\parindent0pt
where the subscripts refer to the interacting particle pair. $\varepsilon_{\rm bb}>0$ and $\varepsilon_{\rm ss}>0$ represent repulsion between pairs of b and s particles, respectively, corresponding to energy penalty for overlap due to entropic effects. $\varepsilon_{\rm bs}$ imposes an attraction between b and s, whereas $r$ is the distance between the centers of the particles. In water solutions, such effective attraction typically rises between hydrophobic components as a result of entropic contributions. It could also be generated by charge distributions on the molecules. Specifically, for charged species such as polyelectrolytes, this interaction can be easily tuned by salt. A long tail repulsion via $\varepsilon_{\rm bs}^{+} > 0$ is useful in preventing singularities. $R_{\rm bb}$ and $R_{\rm ss}$ define the sizes of the particles (typically comparable to the radius of gyration for polymers). Similar theoretical approaches have been used to describe other forms of self-assembly, such as superlattice structures \cite{somerville2018density} and quasicrystals \cite{scacchi2020quasicrystal}. 

In order to choose a relevant set of parameters for Eq. (\ref{pairpot}), we consider the linear dispersion relation branches $\omega_{\pm}(k)$, derived in Ref. \cite{AndyWaltersThieleKnobloch}, which characterise the growth or decay of density modulations, with wave number $k$, in the liquid state. The dependence of this quantity with the different parameters in Eq.~(\ref{pairpot}) is rather complicated. Before investigating such a complex quantity, we start by assuming that the mixture is formed by two independent components. This can be ensured by carefully selecting the form of (\ref{pairpot}). In order to have two clearly distinct length scales we set the ratio $k_{\rm s}/k_{\rm b}\approx 20$, where $k_{\rm s}$ and $k_{\rm b}$ correspond to the large and small particle characteristic wave lengths, respectively. We set $R_{\rm bb}=5$ and $R_{\rm ss}=0.25$, and use dimensionless units throughout. When considering two separate systems of b and s particles, respectively, the linear dispersion relation for each component reads \cite{ArcherEvans, archer2006dynamical, archer2013quasicrystalline, archer2015soft, walters2018structural}
\begin{equation}
\omega_i(k)=-\frac{k^2D_i}{S_i(k)}, \quad\quad i={\rm b},{\rm s};\label{dispersion}
\end{equation}
where $D_i$ is the diffusion coefficient and $S_i(k)$ the liquid structure factor of component $i$, which can be measured in experiments, and is accessible through $S_i(k)=1/(1-\rho_i\hat{c}_i(k))$ \cite{evans_92}, where $\hat{c}_i(k)$ is the Fourier transform of the direct pair correlation function (see SM for details) and $\rho_i$ is the bulk density of species $i$. 
With this approach, one can easily obtain the values of $\varepsilon_{ii}$ necessary to achieve linear instability for a given critical value of densities $\rho_i^{\rm c}$, $i={\rm b,s}$, and vice versa. It is convenient (but not mandatory) to set $\rho_{\rm b}\approx \rho_{\rm s}$. We choose $\beta\varepsilon_{\rm bb}=0.45$ and $\beta\varepsilon_{\rm ss}=70$, where $\beta=1/(k_{\rm B} T)$, $k_{\rm B}$ being the Boltzmann constant and $T$ the temperature. This choice implies that under the hypothesis of two independent particle systems, the critical densities are $\rho_{\rm b}^{\rm c}\approx 0.55$ and $\rho_{\rm s}^{\rm c}\approx 0.78$. These values are obtained by finding the corresponding bulk densities equivalent to the onset of linear instability in Eq.~(\ref{dispersion}), i.e.  $\omega_i(k)=0$ and ${d\omega_i(k)}/{dk}=0$. A common choice for the cross interaction radius is $R_{\rm bs}=\frac{1}{2}(R_{\rm bb}+R_{\rm ss})\approx 3$. In order to avoid a singularity in the density distribution, one should restrict to $2\pi\int_0^{\infty} dr r \phi_{\rm bs}(r) > 0$. However, negative values of the integrated strength can be compensated by the strength of repulsion arising from $\phi_{\rm bb}(r)$ and $\phi_{\rm ss}(r)$. Furthermore, the attractive part of $\phi_{\rm bs}(r)$ should be strong enough to favor mixing, and the repulsive part must be small enough to avoid phase separation. Here we choose $\beta\varepsilon_{\rm bs}=-0.45$ and $\beta\varepsilon_{\rm bs}^{+} = 0.02$.  
Designing Eq.~(\ref{pairpot}) with the conditions provided above allows us to have a mixture in which $\rho_{i}^c$ does not influence $\rho_{j}^c$, for $i\neq j$. This property is evident in the fact that the instability lines in the phase diagram of Fig.~ \ref{diagram} (to be explained below) are almost vertical and horizontal, respectively. This condition is not required; however, it is very helpful in avoiding complications due to the interplay of $\phi_{\rm bs}$ in $\omega_{\pm}(k)$. In fact, under some conditions this interplay could, e.g., partially or fully suppress one or both instabilities. The model parameters used in Eq.~(\ref{pairpot}) are summarized in Table 1.
\begin{table}
\centering
\begin{tabular}{| c | c | c | c | c | c | c | c|}\hline
$\beta\varepsilon_{\rm bb}$  & $\beta\varepsilon_{\rm bs}$ & $\beta\varepsilon_{\rm bs}^{+}$ & $\beta\varepsilon_{\rm ss}$  & $R_{\rm bb}$ & $R_{\rm bs}$ & $R_{\rm ss}$\\  \hline\hline
0.45  & -0.45 & 0.02 & 70  & 5 & 3 &  0.25  \\   \hline
\end{tabular}\caption{Set of dimensionless parameters used in Eq.~(\ref{pairpot}).}\label{table1}
\end{table}
We emphasize that results similar to those obtained here are expected for different parameter sets, but also for models different from Eq.~(\ref{pairpot}).
\begin{figure}
\includegraphics[width=1.0\linewidth]{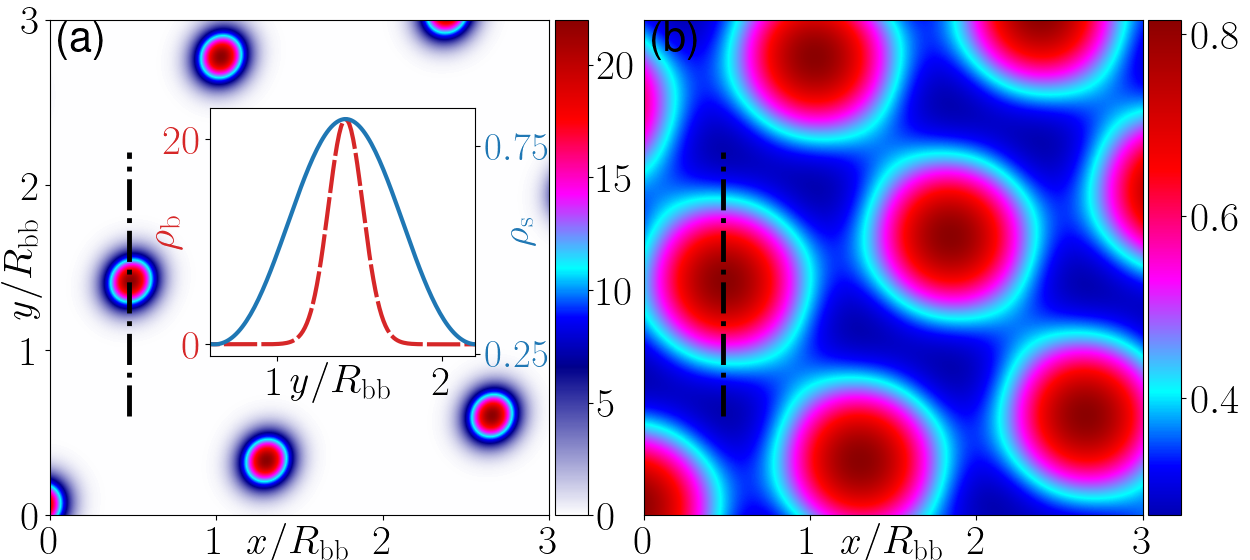}\caption{Single-length-scale self-assembled phase (large dot in Fig.~\ref{diagram}, $\rho_{\rm b}=0.9$, $\rho_{\rm s}=0.45$). This corresponds to e.g. core-shell micelles in  polymer systems or a regular emulsion where droplets form into the carrier phase. (a) shows $\rho_{\rm b}(\textbf{r})$, (b) $\rho_{\rm s}(\textbf{r})$, and the inset the density profiles calculated along the dash-dotted lines at $x=0.476R_{\rm bb}$. The solid line represents $\rho_{\rm s}(y)$ and the dashed line the (much larger) density $\rho_{\rm b}(y)$. The system size is $6.22\times 6.22\>R_{\rm bb}^2$.}\label{circle}
\end{figure}
%
 The linear dispersion relation contains all the information required to determine whether or not the uniform liquid is stable with respect to any (small) perturbation. Thus, to resolve thermodynamically stable states where also assemblies with multiple length scales emerge, we address $\omega_{\pm}(k)$ and look for the crossing point of the two linear instability lines (the dashed and dashed-dotted almost horizontal and vertical lines in Fig.~\ref{diagram}). We could have used also Eq.~(\ref{dispersion}) to approximate the crossing point of the two linear instability lines, but $\omega_{\pm}(k)$ provides the exact information. Unfortunately, knowing this crossing point merely suggests the values of $\rho_{\rm b}$ and $\rho_{\rm s}$ that potentially exhibit equilibrium multiple-length-scale structure. 

The full assembly phase diagram, Fig.~\ref{diagram}, based on Eq. (\ref{pairpot}) and the set of parameters in the Table~\ref{table1}, is obtained by varying $\rho_{\rm b}$ and $\rho_{\rm s}$. An extensive description of this process is reported in the SM. To facilitate comparison with computer simulations on multicomponent polymer systems in a solvent, we adopt for description of the phases terminology that is specific to polymer self-assembly.
The diagram consists of four distinct phases: (i) a phase where both densities are uniform (squares); (ii) a core-shell (or single-core) micelle phase (dots); (iii) a multi-length-scale phase (stars), composed by two subsets (iiia) and (iiib); and (iv) a phase in which the s particles form a hexagonal structure across the system while the b particles are homogeneously spread (rhombi). The boundary lines between the different states in Fig.~\ref{diagram} are meant as a guide for the eye, and should not be considered as exact.

We start the description of the different phases with subset (iiia), commensurate with multi-core-micelle formation in polymer assemblies. An example of this structure is shown in Fig.~\ref{star} (large star in Fig.~\ref{diagram}). Here the b particles form a hexagonal structure with lattice spacing of $\approx R_{\rm bb}$, as shown in Fig.~\ref{star}(a). Simultaneously, the s particles form islands of small hexagonal clusters with lattice spacing of $\approx R_{\rm ss}$, centered at the density maxima of b (panel (b)). Importantly, the orientations of the s islands are independent of each other at each b maximum. Thus, the s particles are locally ordered within each b maximum only. Such assembly response describes complex ordered phases in soft matter such as multi-core micelles or multiple-emulsions. Note that the perfect hexagonal order is due to periodic boundary conditions and lack of thermal fluctuations in the present calculations.

The structural change from (iiia) to (iiib) consists of the s particles islands increasing in size (Fig.~\ref{star}(b)). Increasing $\rho_{\rm s}$ eventually bridges the islands and makes them indiscernible. As the s islands differ in orientation, grain boundaries will emerge. We mark the states with at least two merged islands with orange stars. The extreme case is a structure where the s particles fill completely the space (large values of $\rho_{\rm s}$). 

In contrast, decreasing $\rho_{\rm s}$ starting from states in (iiia) causes the hexagonal arrangement of small islands s to melt into large s droplets centered around the maxima of $\rho_{\rm b}(\textbf{r})$. However, the attractive nature of $\phi_{\rm bs}$ forces the s particles to remain in the vicinity of such maxima (Fig.~\ref{circle}(b) and inset of Fig.~\ref{circle}(a) show an example corresponding to the large dot in Fig.~\ref{diagram}). This transition corresponds to moving from phase (iii) to phase (ii). The latter structure is commensurate with single-core (core-shell) micelles. These states survive for decreasing $\rho_{\rm s}$ unless $\rho_{\rm b} < \rho_{\rm b}^{c}$. Figure~\ref{diagram} shows that multi-core micelles are obtained mostly for $\rho_{\rm s} < \rho_{\rm s}^{\rm c}$, i.e. for densities at which the system is linearly stable with respect to $k_{\rm s}$. The deviation of the actual phase boundaries from the linear instability line can be explained as follows; consider a single-core-micelle state: if the average density of the s particles in each of these peaks is roughly of the same order as $\rho_{\rm s}^{\rm c}$, the s particles find energetically favourable to form clusters with typical spacing of $\approx 2\pi/k_{\rm s}$ ((ii) $\rightarrow$ (iii)). This can be seen as a local instability occurring at the level of a single b site.
 
The two purple stars in (iiia) represent a state in which most of the s clusters behave as in the case represented by green stars (Fig.~\ref{star}(b)). However, some random clusters of s particles do not display the short length scale order but instead their distribution is similar to the one of the b particles (Gaussian-like distributed) (Fig.~\ref{star}(a)). These states might be metastable due to the vicinity with phase boundaries. 

In (iv), the system is filled by a hexagonal structure with spacing of $\approx R_{\rm ss}$. Simultaneously, $\rho_{\rm b}(\textbf{r})$ is uniform. This phase can be found for $\rho_{\rm s}>\rho_{\rm s}^{\rm c}$ and $\rho_{\rm b}<\rho_{\rm b}^{\rm c}$. The attractive nature of $\phi_{\rm bs}$ favours the orange-star states to exist down to values of $\rho_{\rm b}<\rho_{\rm b}^{\rm c}$, where one would expect to find rhombi states if only linear instabilities were considered. Cluster formation of b particles is enhanced by the attraction mediated by clusters of s particles. This can be explained in a manner similar to the discrepancy between the linear instability line of the s particles and the boundary between dots and stars.
\begin{table}
\centering
\begin{tabular}{ | c || c | c | c | c |}\hline
State & i & ii & iii & iv \\ 
\hline\hline
$\rho_{\rm b}$ & even & hex $R_{\rm bb}$ & hex $R_{\rm bb}$  & even \\ 
\hline
$\rho_{\rm s}$ & even & hex $R_{\rm bb}$ & hex $R_{\rm ss}$ islands over hex $R_{\rm bb}$ & hex $R_{\rm ss}$  \\ 
\hline
\end{tabular}\caption{Symmetries of $\rho_{\rm b}$ and $\rho_{\rm s}$ in the different phases.}\label{table2}
\end{table}

Finally, in phase (i) both species are uniformly distributed, i.e. a solution of miscible species. The slight discrepancy between the linear instability line of particles b and the boundary between the homogeneous state and the states in which the latter species is linearly unstable is due to finite-size effects and to the fact that the dispersion relation only considers the linear contributions of the dynamics. Furthermore, the size of the system has been chosen randomly which may lead to a wave number slightly different from $k_{\rm b}$. $\omega_{+}(k_{\rm b})$ is concave with a maximum at $k_{\rm b}$ at the onset of the instability: particles b will form clusters for a slightly different value of $\rho_{\rm b}$. 
The symmetries of $\rho_{\rm b}(\textbf{r})$ and $\rho_{\rm s}(\textbf{r})$ in the different phases are summarized in Table~\ref{table2}. We note that the transition between phases (ii) and (iii) is discontinuous. This is due to spatial symmetry breaking between these phases. 

To compare the emergence of the above-discussed single vs. multiple-length-scale assembly in a realistic molecular level system, we perform dissipative particle dynamics (DPD)~\cite{hoogerbrugge1992simulating,espanol1995statistical, groot1997dissipative,espanol2017} simulations of a polymeric solution composed of a solvophobic polymer (A$_{19}$) and a di-block copolymer (A$_1$B$_6$) in a solvent using the LAMMPS~\cite{plimpton1995} simulation package. The segments A and B are solvophobic and solvophilic, respectively. The subscripts refer to the number of DPD beads (block lengths). Each bead represents a group of atoms which experience a force $\textbf{F}_{i}=\sum_{j\neq i}\left[\textbf{F}^{\rm C}_{ij}+\textbf{F}^{\rm D}_{ij}+\textbf{F}^{\rm R}_{ij}\right]$, $i,j=1, \cdots, N$, where $\textbf{F}^{\rm C}$ describes conservative interactions, $\textbf{F}^{\rm D}$ dissipative contributions, $\textbf{F}^{\rm R}$ random contribution and where $N$ is the total number of beads. The interaction forces are treated as pairwise additive and are truncated at a distance $r_c$. In the polymer chains, adjacent beads also contribute to a spring-like force $\textbf{F}^{\rm S}$. These soft potentials facilitate acceleration of the numerical simulations so that realistic experimental time and length scales can be achieved. Further details and simulation specifics can be found in the SM. 

\begin{figure}[t!]
\centering\includegraphics[width=1\linewidth]{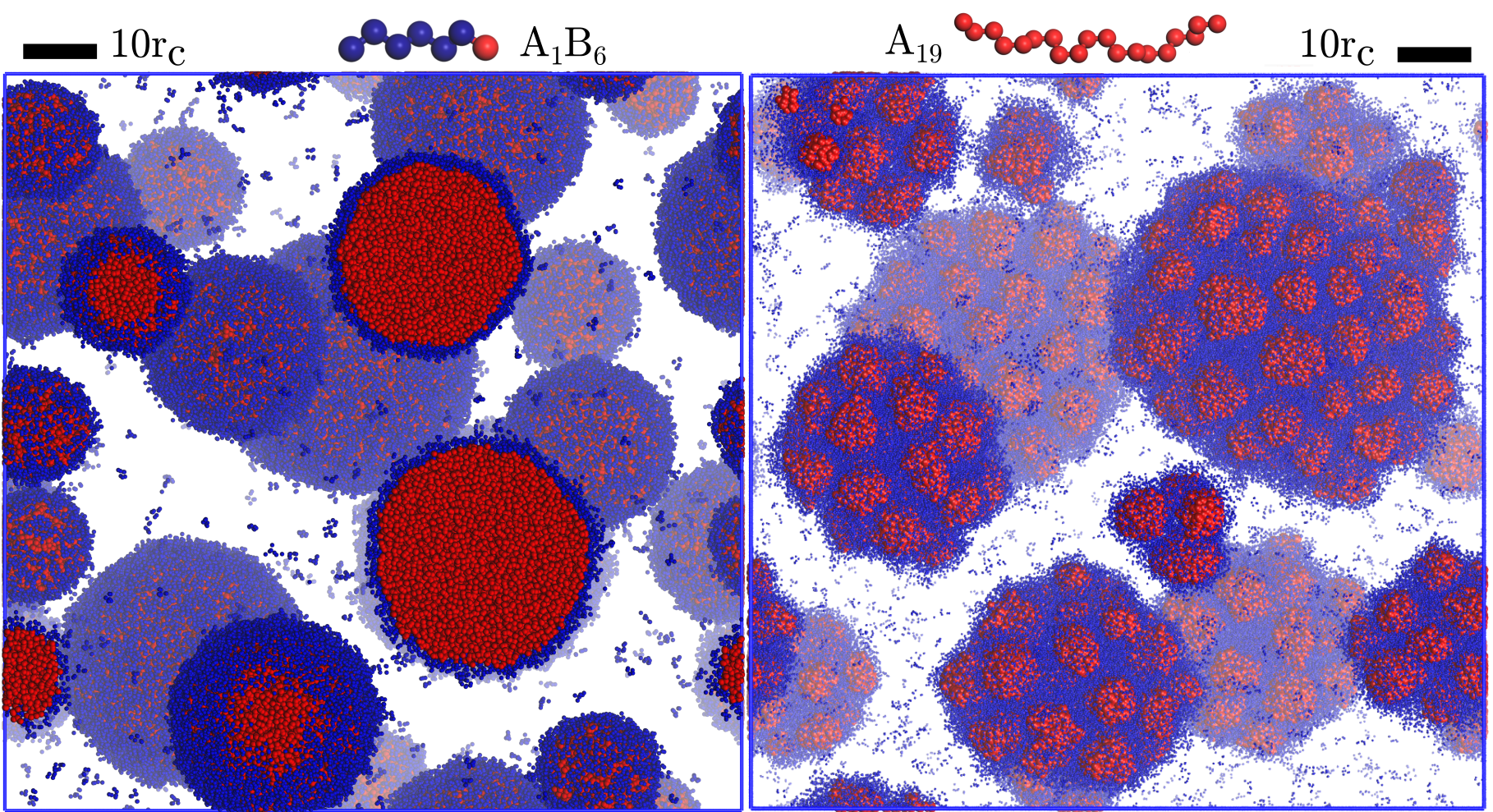}\caption{Equilibrium configurations from 3D DPD simulations of a polymeric solution composed of a solvophobic polymer (A$_{19}$) and a di-block copolymer (A$_1$B$_6$) in a solvent. The segments A (red) and B (blue) are solvophobic and solvophilic, respectively. The subscript refers to the number of DPD beads. Left: single-core (core-shell) configuration for molar fractions 50\% A$_{19}$ and 50\% A$_1$B$_6$. Right: multi-core configuration for molar fractions 10\% A$_{19}$ and 90\% A$_1$B$_6$. The solvent is not shown for clarity. 
}\label{simulations}
\end{figure}

Figure~\ref{simulations} shows snapshots of different equilibrium configurations obtained from simulations by varying the molar fractions of the components. This system strikingly self-assembles into structures corresponding to a multi-core-corona and a single-core-corona states. Both cases correspond to a total solid content of $N_{\rm s}/(N_{\rm s}+N_{\rm w})=20.15$\% in aqueous solvent, where $N_{\rm s}$ and $N_{\rm w}$ are the number of solid and water beads, respectively. In the former case the molar fractions are $10$~\% A$_{19}$ and $90$~\% A$_1$B$_6$, whilst in the latter $50$~\% A$_{19}$ and $50$~\% A$_1$B$_6$. The simulations are performed in a cubic box of $100 \times 100 \times 100\ r_c^3$.
These equilibrium configurations show dual and single-length-scale structural assemblies analogous to those in our DFT based phase diagram (phases (ii) and (iiia) in Fig. \ref{diagram}). It is tempting to interpret a DPD polymer chain as a single coarse-grained DFT particle, but it should be noted that the surfactant-like nature of the copolymer makes a quantitative comparison between chemically predictive molecular model and DFT-assembly structures difficult. Nevertheless, the two distinct length scales emerging from the effective interactions in Fig.~\ref{star} also appear in the DPD model. Additionally, also in DPD simulations the key feature for multi-core assembly is a sufficient degree of immiscibility between the species. The formation of solvophobic cores needs to be energetically favorable and need to be sufficiently stabilized by the solvophilic polymer segments, and the solvophilic polymer cannot have too favorable interactions with either the solvent or the solvophobic polymer. If either the solvophobic polymers are too solvophobic, or the solvophilic one too solvophilic, or the two polymers readily mix, the assembly becomes core-shell or phase separates. This demonstrates how competition between interactions and mutual balance of immiscibility gives rise to multiple-length-scale assemblies -- single-length-scale self-assembly is retained when any of the pairwise effective interactions dominates. 

%
%
To summarize, we have presented a microscopic theory capable of explaining self-assembly within soft matter with multiple competing length scales. The theory relies on the interplay between the effective interaction potentials modelling the different constituents of the system. 
The significance of our findings resides in the identification of key molecular features involved in general multiple-length-scale self-assembly. Although multiple-length-scale assembly is ubiquitous, such guidelines have been lacking until now. By providing much needed insight to understanding, e.g., morphologies rising in intracellular biocondensates, multiple emulsions, or lipid bilayer microdomains, we obtain means to engineer and tune soft matter to desired multiple 
-length-scale structures. In addition to obvious applications to multiple-length-scale self-assembling compartmentalization, such as drug delivery, catalysis and selective multistep reaction platforms, our work provides tools to harnessing the full potential of revolutionary materials production via biological mechanisms, advanced engineered biomaterials, or complex polymer assemblies, providing crucial insight on how to tune the assembly response.\\\\
{\it Acknowledgements}: A.S. is thankful to Andrew J. Archer for fruitful discussions. This work was supported by Academy of Finland grants No. 309324 (M.S.) and Nos. 307806 and 312298 (T.A-N.). T.A-N. has also been supported by a Technology Industries of Finland Centennial Foundation TT2020 grant. We are grateful for the support by FinnCERES Materials Bioeconomy Ecosystem. Computational resources by CSC IT Centre for Finland and RAMI -- RawMatters Finland Infrastructure are also gratefully acknowledged.\\\\
{\it Author contributions}: This work was conceived and supervised by T.A-N. and M.S. A.S. formulated and executed the theoretical calculations and wrote the first draft of the manuscript. S.J.N. carried out the molecular simulations. All authors contributed to writing.

\bibliographystyle{ieeetr}
\bibliography{main.bib}
\onecolumngrid
\vspace{115mm}

\begin{center}
{\Large \bf Supplementary Material}
\end{center}

\setcounter{equation}{0}
\renewcommand{\theequation}{S\arabic{equation}}

\subsection*{Density functional theory for mixtures and the Ramakrishnan and Yussouff approximation}
The grand potential of a system composed of two types of particles is %
\begin{equation}
\Omega[\rho_1,\rho_2]=\mathcal{F}[\rho_1,\rho_2]+\sum_{i=1,2}\int d\textbf{r}\left[ V_i^{\rm ext}(\textbf{r})-\mu_i\right]\rho_i(\textbf{r}),\label{grand_canonical}
\end{equation}
where $\mathcal{F}$ is the intrinsic Helmholtz free energy functional, $\rho_1=\rho_1(\textbf{r})$ and $\rho_2=\rho_2(\textbf{r})$ are the spatially varying densities of the two species, $V_i^{\rm ext}(\textbf{r})$ is the one-body external potential acting on species
$i$ (for bulk systems $V_i^{\rm ext}(\textbf{r}) \equiv 0$ for $i = 1, 2$) and $\mu_i$ are the chemical potentials. The intrinsic Helmholtz free energy can be split into two terms
\begin{equation}
\mathcal{F}[\rho_1,\rho_2]=\mathcal{F}^{\rm id}[\rho_1,\rho_2]+\mathcal{F}^{\rm exc}[\rho_1,\rho_2],\label{free_energy}
\end{equation}
where the first term is the ideal gas contribution, i.e. 
\begin{equation}
\mathcal{F}^{\rm id}[\rho_1,\rho_2]=k_{\rm B} T \sum_{i=1,2}\int d\textbf{r} \rho_i(\textbf{r})\left[\ln\left(\Lambda_i^{\rm d}  \rho_i(\textbf{r})\right)-1\right],\label{ideal}
\end{equation}
and $\Lambda_i$ is the thermal de Broglie wavelength, ${\rm d}$ is the dimensionality of the system, $T$ the temperature and $k_{\rm B}$ the Boltzmann constant. The second term in Eq. (\ref{free_energy}) is the excess Helmholtz free energy arising from the interactions between the particles. Following Ramakrishnan and Yussouff \cite{ramakrishnan1979first}, we approximate this functional by a functional Taylor expansion around the homogeneous fluid states $\rho_{0,i}$. A truncation of the series expansion at second order gives 
\begin{equation}
\mathcal{F}^{\rm exc}[\rho_1,\rho_2]=\mathcal{F}^{\rm exc}[\rho_{0,1},\rho_{0,2}]+\sum_{i=1,2}\int d\textbf{r}\mu_i^{\rm exc}\delta\rho_i(\textbf{r})-\frac{1}{2\beta}\sum_{\substack{i=1,2 \\ j=1,2}}\int d\textbf{r}\int d\textbf{r}'\delta\rho_i(\textbf{r})c_{ij}(\mid \textbf{r} -\textbf{r}'\mid)\delta\rho_j(\textbf{r}'),\label{excess}
\end{equation}
where $\delta\rho_i(\textbf{r}) = \rho_i(\textbf{r}) - \rho_{0,i}$ and $\mu_i^{\rm exc} = \mu_i - k_{\rm B} T \ln(\rho_{0,i}\Lambda_i^{\rm d})$ are the excess chemical potentials. The pair direct correlation functions $c_{ij}(\textbf{r})$ are obtained via the random phase approximation (RPA). The RPA consists of assuming a simple mean-field form for the excess free energy in Eq.~(\ref{free_energy}). This leads to $c_{ij}(\textbf{r})=-\beta \phi_{ij}(\textbf{r})$, where $\phi_{ij}(\textbf{r})$ are the effective pair potentials, and $\beta=(k_{\rm B}T)^{-1}$~\cite{likos2001effective}. For additional accuracy, one could, for example, use the hypernetted chain (HNC) Ornstein-Zernike integral equation theory \cite{hansen_mcdonald}.
The equilibrium density profiles $\rho_i(\textbf{r})$ are those which minimise the functional of the grand potential $\Omega[\rho_1,\rho_2]$ and which therefore satisfy the following pair of coupled Euler-Lagrange equations
\begin{equation}
\frac{\delta \Omega[\rho_1,\rho_2]}{\delta\rho_i}=0,
\end{equation}
for $i=1,2$.

\subsection*{Difference in the grand canonical potential energy}
Substituting Eqs.~(\ref{free_energy}-\ref{excess}) into Eq.~(\ref{grand_canonical}) one obtains an expression for the grand canonical potential. For homogeneous system, i.e. $\rho_1(\textbf{r})=\rho_{0,1}$ and $\rho_2(\textbf{r})=\rho_{0,2}$, the latter reduces to
\begin{equation}
\Omega^0[\rho_{0,1},\rho_{0,2}]=\frac{1}{\beta}\sum_{i=1}^2\int d\textbf{r} \rho_{0,i}\left[\ln\left(\Lambda^{\rm d}\rho_{0,i}\right)-1-\beta\mu_i\right]+\mathcal{F}^{\rm exc}[\rho_{0,1},\rho_{0,2}].\label{grand_canonical_0}
\end{equation}
The difference in the grand canonical potential energy between the equilibrium state and the corresponding homogeneous state becomes
\begin{equation}
\begin{split}
\Delta\Omega=\Omega[\rho_{1},\rho_{2}]-\Omega^0[\rho_{0,1},\rho_{0,2}]&=\frac{1}{\beta}\sum_{i=1}^2\int d\textbf{r}\left[\rho_i(\textbf{r})\ln\left[\frac{\rho_i(\textbf{r})}{\rho_{0,i}}\right]-\delta\rho_i(\textbf{r})\right]\\
&-\frac{1}{2\beta}\sum_{\substack{i=1,2 \\ j=1,2}}\int d\textbf{r}\int d\textbf{r}'\delta\rho_i(\textbf{r})c_{ij}(\mid \textbf{r} -\textbf{r}'\mid)\delta\rho_j(\textbf{r}').\label{delta_grand_canonical}
\end{split}
\end{equation}

\subsection*{Dissipative particle dynamics simulations (DPD)}
The DPD simulation method, originally proposed by Hoogerbrugge and Koelman ~\cite{groot1997dissipative}, is a mesoscale coarse-grained bead-based molecular simulation technique. Combining aspects of molecular dynamics and lattice-gas automata, DPD acknowledges the idea that different beads can overlap, modelling non-bonded interactions with soft repulsive potentials~\cite{groot1997dissipative,espanol2017}. The soft potentials employed to describe the interactions between the beads allow the simulations to reach realistic experimental time and length scales for e.g. block-copolymer self-assembly systems. Each DPD bead represents a coarse-grained region in the molecular system (e.g. several monomers of a polymer, or a solvent region, or a group of atoms) which experience a force 
\begin{equation}
\textbf{F}_{i}=\sum_{j\neq i}\left(\textbf{F}^{\rm C}_{ij}+\textbf{F}^{\rm D}_{ij}+\textbf{F}^{\rm R}_{ij}\right),\quad i,j=1, \cdots, N,
\end{equation}
where $\textbf{F}^{\rm C}$ describes conservative interactions, $\textbf{F}^{\rm D}$ dissipative contributions, $\textbf{F}^{\rm R}$ a random contribution, and $N$ corresponds to the number of DPD beads in the system. The interaction forces are treated as pairwise additive and are truncated at a distance $r_{\rm c}$. 

The conservative force is a soft repulsive force acting along the centers of two DPD particles, given by
\begin{equation}
\textbf{F}^{\rm C}_{ij}=a_{ij}(1-r_{ij})\hat{\textbf{r}}_{ij},
\end{equation}
where $a_{ij}$ is the maximum repulsion between beads $i$ and $j$, $r_{ij}=\mid r_i - r_j\mid/r_{\rm c}$ is the cut-off normalized distance between beads $i$ and $j$, and   $\hat{\textbf{r}}_{ij}=\textbf{r}_{ij}/r_{ij}$ gives the force direction via a unit vector. The coefficient $a_{ij}$ is connected to the Flory-Huggins mixing parameter $\chi_{ij}$  via the relation $\chi_{ij}=(a_{ij}-a_{ii})/3.27$ at a density $\rho=3$. The choice $a_{ii} = 25$ for the repulsion parameter between beads of the same species (i.e., $\chi_{ii}=0$) is common and based on the compressibility of the dilute solution~\cite{groot1997dissipative}. An $a_{ij}$ value exceeding the self-repulsion in magnitude corresponds to a stronger bead-bead repulsion between beads of different species. 

The dissipative force is given by
\begin{equation}
\textbf{F}_{ij}^{\rm D}=-\gamma\omega^{\rm D}(r_{ij})(\hat{\textbf{r}}_{ij}\cdot\textbf{v}_{ij})\hat{\textbf{r}}_{ij},
\end{equation} 
where $\gamma$ is a viscosity related parameter ($\gamma = 4.5$), $\omega^{\rm D}$ is a weight function that goes to zero at $r_{\rm c}$, and the relative velocity is $\textbf{v}_{ij} = \textbf{v}_i -\textbf{v}_j$. 

The random force is given by
\begin{equation}
\textbf{F}^{\rm R}_{ij}=\sigma\omega^{\rm R}(r_{ij})\xi_{ij}\hat{\textbf{r}}_{ij},
\end{equation}
where $\xi_{ij}$ is a zero-mean Gaussian random variable of unit variance and $\sigma^2 = 2\gamma k_{\rm B} T$.
The weight functions follow the relation $\omega^{\rm D}(r_{ij}) = \omega^{\rm R}(r_{ij})^2 = (1-r_{ij})^2$ for $r_{ij} < r_{c}$. Consecutive beads in a polymer chain also perceive a spring force $\textbf{F}^{S}_i$ defined by
\begin{equation}
\textbf{F}^{\rm S}_i=-\kappa\sum_{j^{*}} (r_{ij}-r_0)\hat{\textbf{r}}_{ij},
\end{equation}
where $\kappa$ is the spring constant, which is set to $\kappa=80$ in this work, the equilibrium distance $r_0=r_c$, and where $j^{*}$ refers to the nearest neighbours in the chain. 

The system modelled in this work is composed of a mixture of a solvophobic polymer (${\rm A}_{19}$) and a di-block  copolymer (${\rm A}_1{\rm B}_6$) in a solvent (${\rm S}$) medium. The nomenclature ${\rm A}$, ${\rm B}$, ${\rm S}$ refers to the DPD beads such that ${\rm A}$ is a solvophobic bead, ${\rm B}$ a solvophilic bead and ${\rm S}$ the solvent bead. The subscripts refer to the number of beads in the chain. For the interactions between beads of same type, we use $a_{\rm AA} = a_{\rm BB} = a_{\rm SS} = 25$. This value is based on the compressibility of the dilute solution~\cite{groot1997dissipative}. The interactions between the beads in the simulations are  $a_{\rm AB} = 72$, $a_{\rm AS} = 115$ and $a_{\rm BS} =30$.
To simplify the DPD equations and simulations, the cutoff radius $r_{\rm c}$, the bead mass $m$, and the energy $k_{\rm B}T$ are reduced to $r_{\rm c} = m = k_{\rm B}T = 1$ which leads to the time unit $\tau = (mr_{\rm c}^2/ k_{\rm B}T)^{1/2} = 1$.

The DPD simulations were performed using the LAMMPS ~\cite{plimpton1995} package. A modified version of the velocity-Verlet algorithm is used to integrate the equations of motion. A time step of $\Delta t = 0.05\tau$ is used.  The simulations are performed in a cubic box of $100 \times 100 \times 100\ r_{\rm c}
^3$ in size. Periodic boundary conditions in all directions in 3D were used. The system is initialized with setting the polymers and the solvent beads with random placement in the box (overlap excluded in initialization). The simulations were carried out for $2 \times 10^6$ time steps, and equilibration checked for via analysis of the time evolution of the assembly structures. 

\end{document}